\documentclass[reprint,superscriptaddress,nobibnotes,amsmath,amssymb,twocolumn,prb]{revtex4-2}
\usepackage[pdftex]{graphicx}
\usepackage[english]{babel}
\usepackage{physics}
\usepackage{tabularx}
\usepackage{float}
\usepackage{amssymb}
\usepackage{amsmath}
\usepackage{dsfont}
\usepackage{epsfig}
\usepackage{dcolumn}
\usepackage{bm}
\usepackage{amstext}
\usepackage{mdframed}
\usepackage{siunitx}

\newcommand{\be}{\begin{equation}}
\newcommand{\ee}{\end{equation}}
\newcommand{\FN}{\text{FN}}
\newcommand{\ex}{\text{ex}}

\begin{document}
\title{Self-Healing Diffusion Monte Carlo applied to
    a simple fermionic model: A critical assessment of the method}
\author{Michel Caffarel}
\thanks{Corresponding author: caffarel@irsamc.ups-tlse.fr}
\affiliation{Laboratoire de Chimie et Physique Quantiques (UMR 5626), CNRS and Universit\'e de Toulouse, France}
\author{Manon Pinar}
\affiliation{Laboratoire de Chimie et Physique Quantiques (UMR 5626), CNRS and Universit\'e de Toulouse, France}
\author{Anthony Scemama}
\affiliation{Laboratoire de Chimie et Physique Quantiques (UMR 5626), CNRS and Universit\'e de Toulouse, France}

\begin{abstract}
We investigate the Self-Healing Diffusion Monte Carlo (SHDMC) method using a one-dimensional model
with periodic boundary conditions.
An inversion symmetry is introduced to mimic the antisymmetry property of
fermionic wave functions, with the bosonic and fermionic sectors being modeled by the even and odd eigenstates, respectively.
As in realistic fermionic systems, the nodal structure is only partially
constrained by symmetry, making this model a non-trivial testbed for nodal optimization algorithms such as SHDMC.
We show that the nodal evolution under SHDMC iterations can be cast into a dynamical system exhibiting
both attractive and repulsive fixed points. When applying SHDMC to this model in the absence of statistical noise,
the fixed-node energy is found to increase during iterations and the node converges to an incorrect value.
This occurs for any finite basis, even when the exact node is representable in it;
the attractive fixed point approaches the exact node only as the basis becomes complete, while the convergence toward it becomes increasingly slow.
We further show that this problem can be largely cured by modifying the nodal update criterion to give more importance
to the nodal region. 
Our results suggest that, in the general case, an efficient nodal optimization may require both an improved nodal update criterion
and a localized basis set allowing local distortions of the trial wave function, as is the case for this model.
\end{abstract}
\noindent
\maketitle
\section{Introduction}
The Self-Healing Diffusion Monte Carlo (SHDMC) method was introduced more than fifteen years ago \cite{Reboredo2009a,Reboredo2009b,Bajdich2010,Reboredo2012,Reboredo2014} as an approach
to improve the nodal surface of fermionic wave functions within the Diffusion Monte Carlo (DMC) framework.
SHDMC is conceptually attractive: It is a simple, iterative scheme in which only fixed-node calculations
are required. At each iteration, the nodal surface is updated using the information obtained from
the preceding fixed-node calculation. This simplicity is in sharp contrast to the commonly employed
strategy of variationally optimizing an approximate trial wave function to get better nodes ({\it e.g.}, Refs. [\onlinecite{Sorella1998,Umrigar2007,Toulouse2008}]).
The efficiency of this approach depends critically both on the quality of the chosen
functional form for the trial wave function
and on the efficiency of the nonlinear optimization of many parameters,
a delicate problem in the presence of statistical noise.
Initial applications of SHDMC have led to significant improvements of the
fixed-node energies of atoms and molecules\cite{Reboredo2009a,Reboredo2009b,Bajdich2010}.
However, the method has seen little use since its introduction.
Very recently \cite{Spanedda2025,Spanedda2026}, new applications to correlated periodic materials
have again produced promising results.
This has motivated us to take a closer look at the theoretical foundations of this approach
and, more specifically, to gain some insight into its general applicability,
beyond specific and possibly favorable examples.
Our aim is not to question the results reported in these applications, but to identify,
in a setting where all quantities can be computed exactly, the conditions under which the
SHDMC iteration converges to the exact node.
In this work, we investigate the SHDMC method on a model system designed to
mimic the more general problem of fermionic systems.

The paper is organized as follows. In Sec. \ref{sec1}, we introduce the simple ``fermionic'' model
used in this work. Section \ref{sec2} presents the SHDMC algorithm in a noise-free context and its application to this model,
showing that, for any finite basis, the method converges to an incorrect node.
In Sec.~\ref{relation}, we discuss the relation between our noise-free implementation and the original
algorithm, and we relate our results to the theoretical arguments of Ref.~\cite{Reboredo2009a}.
In Sec. \ref{sec3}, we introduce a modification of SHDMC designed to put greater emphasis on the nodal region.
We show that this new version of the method does work in a specific regime.
To make this approach suitable for practical DMC calculations, we propose in Sec. \ref{local} the use of a localized basis set,
which, for this model, leads to a well-conditioned scheme.
A summary of the main results is given in Sec. \ref{sec4}.
Finally, the Appendix is devoted to the application of SHDMC to the simple case of a free particle
confined to a finite segment. Although the exact node is known by symmetry and
this system is therefore not representative of realistic fermionic systems, it provides a useful test case
for which the equations can be worked out analytically and the convergence of the algorithm investigated in some
detail.

\section{The fermionic toy model}
\label{sec1}
A simple one-dimensional model that retains the essential nodal properties of a fermionic system
is introduced. In general, one-dimensional systems are not appropriate for this purpose,
because the antisymmetry requirement uniquely fixes the exact nodes of the ground state.
However, this is no longer the case
in the presence of periodic boundary conditions. We therefore consider a single
particle confined to a one-dimensional ring, described by the Hamiltonian
\be
  H = -\frac{1}{2}\frac{d^2}{d\theta^2} + V(\theta),
\ee
whose eigenstates satisfy
\be
  H\psi(\theta) = E\,\psi(\theta),
\ee
with $V(\theta+2\pi)=V(\theta)$ and $\psi(\theta+2\pi)=\psi(\theta)$.

\begin{figure}[t]
\centering
  \includegraphics[width=0.9\columnwidth]{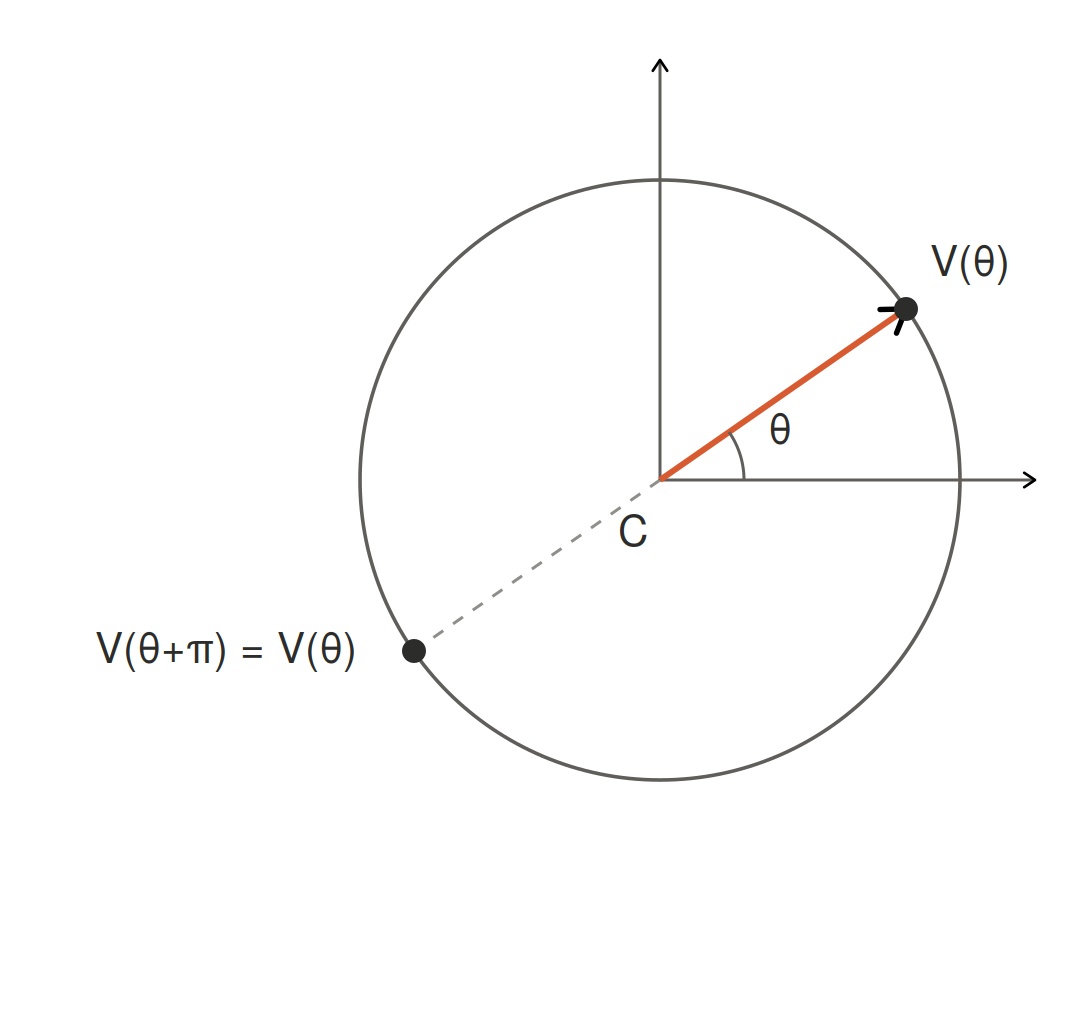}
\caption{The toy model -one particle on a circle- and its inversion symmetry.}
\label{fig1}
\end{figure}

To mimic the Pauli exclusion constraint of fermionic systems, we impose inversion 
symmetry on the potential
\be
  V(\theta+\pi)=V(\theta).
\ee
The system with its inversion symmetry with respect to the center $C$ of the circle is depicted
in Fig.~\ref{fig1}.

As a result of the inversion symmetry, the eigenfunctions can be classified according to their parity
under inversion,
\be
\psi_n(\theta+\pi)=\pm\,\psi_n(\theta), \qquad n=0,1,\ldots
\ee
The even eigenfunctions, including the nodeless ground state $\psi_0$,
constitute the ``bosonic'' sector of the spectrum, whereas the odd
eigenfunctions form the ``fermionic'' sector. The physical fermionic ground
state is the odd-parity eigenstate of lowest energy, namely
$\psi_1(\theta)$.

For this one-dimensional system with periodic boundary conditions, the nodal
set $\mathcal{S}$ of the fermionic ground state consists of two points
on the ring:
\be
\mathcal{S} = \{\theta_{ex}\} \cup \{\theta_{ex}+\pi\}.
\ee
where $\theta_{ex}$ is the exact node.
The important remark is that the nodal structure is not completely
determined by symmetry. Inversion symmetry guarantees that whenever $\theta$
is a node, so is its image $\theta+\pi$. However, the exact value $\theta_{ex}$
is not fixed by symmetry: $\theta_{ex}$ depends on the
specific form of the potential and can only be determined by solving the
Schr\"odinger equation.

In this respect, the model shares this property with realistic
many-electron systems. In those systems, the Pauli principle determines only
the exchange nodes ({\it i.e.}, configurations in which two like-spin electrons
coincide), which define a nodal subset of dimension $3N-3$ for an
$N$-electron system\cite{Klein1976}. The complete nodal hypersurface, however, has dimension
$3N-1$ and its structure depends on the potential.

For the present model, a fixed-node calculation amounts to solving the
Schr\"odinger equation on a single nodal domain, namely the half-ring
$(\theta_{\FN},\,\theta_{\FN}+\pi)$, subject to the Dirichlet boundary conditions
\be
  \psi_\FN(\theta_{\FN}) = \psi_\FN(\theta_{\FN}+\pi) = 0,
\label{dirichlet}
\ee
where $\theta_{\FN}$ denotes the approximate angle used in the fixed-node calculation.
The fixed-node wave function on the complementary half-ring is reconstructed by
antisymmetry under inversion.

To avoid the complications associated with statistical noise, all the fixed-node
calculations of the present work are performed {\it deterministically}. To this end, we expand the
fixed-node solution in a complete orthonormal basis on the nodal domain
$\qty(\theta_{\FN},\theta_{\FN}+\pi)$ that vanishes at the boundaries $\theta_{\FN}$ and
$\theta_{\FN}+\pi$. Specifically, we use
\be
\psi_{\FN}(\theta) = \sum_{k=1}^{\infty} c_k\, w_k(\theta),
\ee
with
\be
  w_k(\theta) = \sqrt{\frac{2}{\pi}}\,\sin\!\bigl[k(\theta-\theta_{\FN})\bigr],
  \qquad k = 1, 2, \ldots
\ee
On the full ring (no fixed-node constraint) a Fourier basis is employed.
More precisely, 
taking into account the inversion symmetry (even harmonics do not contribute)
the exact eigenfunctions are expanded as follows:
\be
\psi(\theta) = \sum_{k=1}^{\infty} \qty[a_k\,u_k\qty(\theta) + b_k\,v_k\qty(\theta)],
\label{Fourier_basis}
\ee
where
\be
u_k(\theta)= \frac{\cos\qty[\qty(2k-1)\theta]}{\sqrt{\pi}}, \qquad
v_k(\theta)= \frac{\sin\qty[\qty(2k-1)\theta]}{\sqrt{\pi}}.
\ee
For simplicity, we denote by $\phi_k$ the basis set functions, whether $u_k$ or $v_k$.

In both basis sets, the Hamiltonian matrix elements
$ \int_{\theta_{\FN}}^{\theta_{\FN}+\pi} w_k(\theta)\,H\,w_l(\theta)\,d\theta$ and
$\int_0^{2\pi} \phi_k(\theta)\,H\,\phi_l(\theta)\,d\theta$ can be evaluated analytically.\\

In this work, numerical results are presented for the potential
\be
  V(\theta) = \cos (2\theta) + \lambda\sin (4\theta)
\ee
with $ \lambda = 1$.\\

\begin{figure}[t]
\centering
  \includegraphics[width=0.9\columnwidth]{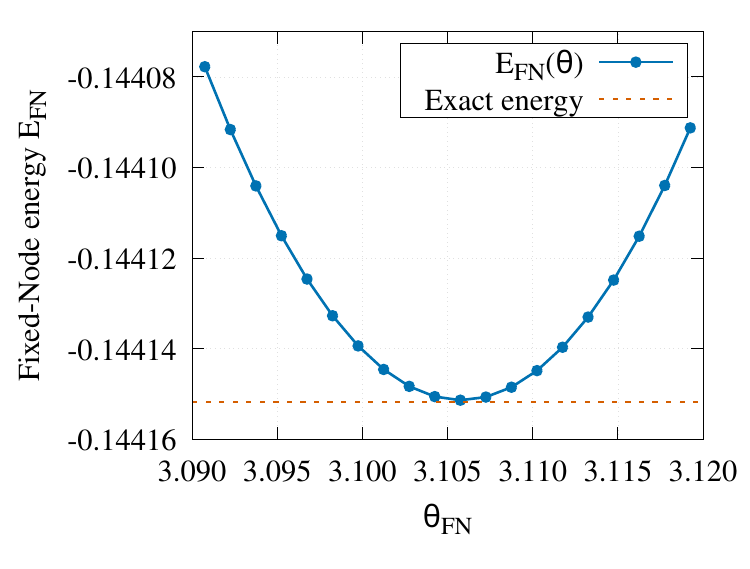}
\caption{Fixed-node energy as a function of the nodal
angle $\theta_{\FN}$ for the model.
The minimum of the fixed-node energy is located at the exact angle, $\theta_{\ex} = \num{3.1058031}...$. The exact energy is
$E_{\ex} = \num{-0.1441519132}... $.}
\label{fig2}
\end{figure}

Figure~\ref{fig2} displays the fixed-node energy as a function of the nodal
position $\theta_{\FN}$; the exact energy is indicated by the horizontal line. As
expected from the variational principle\cite{Reynolds1982},
the fixed-node energy is above the exact energy
and reaches its minimum when the angle equals the exact value. The exact angle
$\theta_{\ex}$ agrees with the node obtained independently
from the exact wave function computed in the $\phi_k$ Fourier basis set. We stress again
that this node (and its image under inversion) cannot be inferred from symmetry alone: as in
realistic fermionic systems, it depends on the specific form of the potential
and requires solving the Schr\"odinger equation.

\begin{figure}[t]
\centering
\includegraphics[width=0.9\columnwidth]{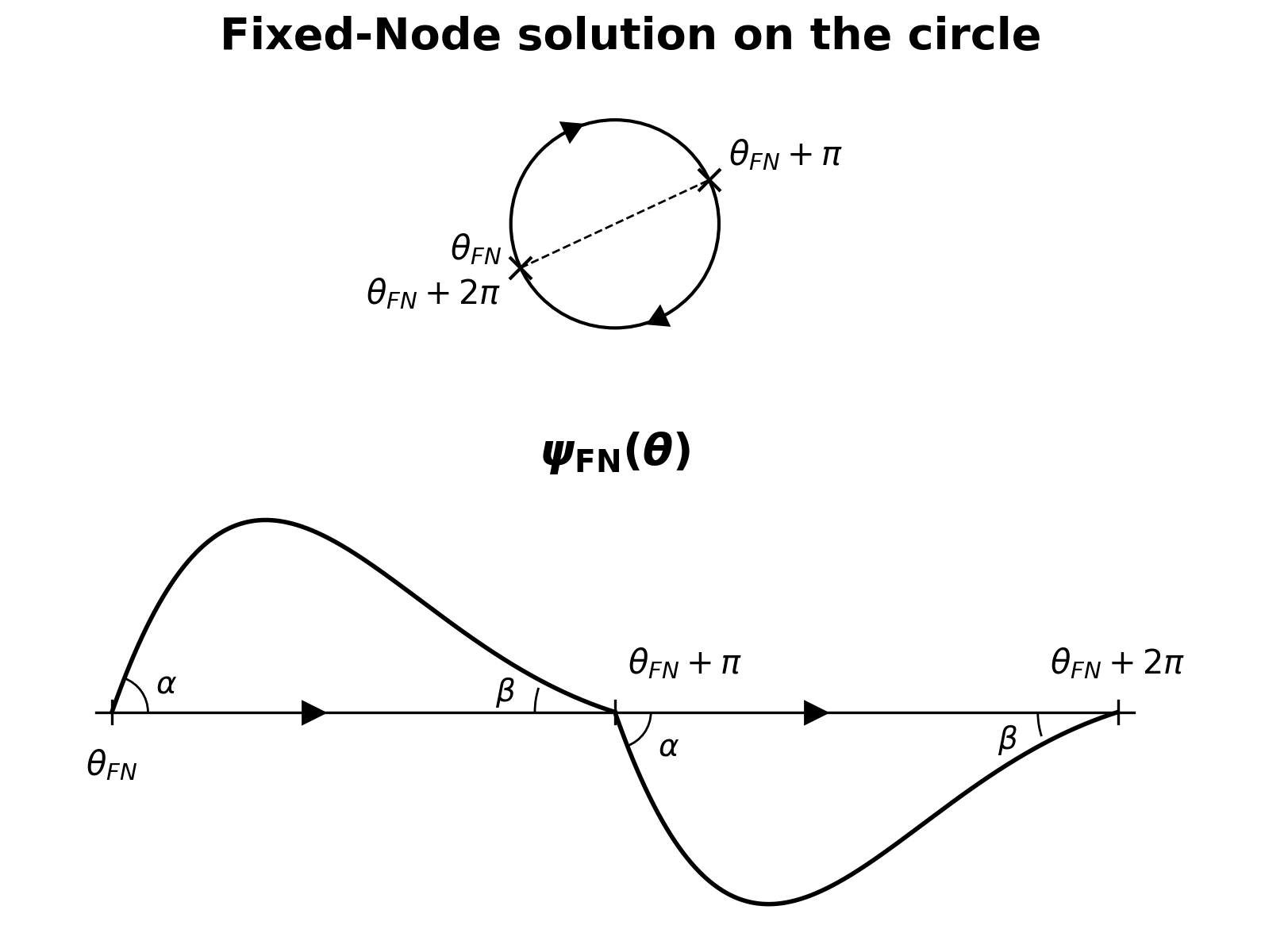}
\caption{
Schematic representation of the fixed-node solution 
$\psi_{\FN}(\theta)$ on the two half-circles delimited by the nodes $\theta_{\FN}$ and $\theta_{\FN}+\pi$.
The angles $\alpha$ and $\beta$
characterize the slopes of $\psi_{\FN}(\theta)$ 
on either side of each node. In general 
$\alpha \ne \beta$, except at the exact nodes where the slopes are identical.
}
\label{fig3}
\end{figure}

A useful diagnostic of nodal quality is provided by the discontinuity of the derivative
of the fixed-node solution at the nodes. Indeed, at an approximate node, the left and
right derivatives of the fixed-node wave function differ and coincide only 
at the exact node.  
The discontinuity of the derivative is illustrated in Fig. \ref{fig3}, where the two pieces of the fixed-node 
wave function, corresponding 
to the two intervals $(\theta_{\FN}, \theta_{\FN}+\pi)$ and $(\theta_{\FN}+\pi,\theta_{\FN}+2\pi)$ are shown.
For the left piece of the FN wave function the slope at the nodes $\theta_{\FN}$ and $\theta_{\FN}+\pi$ are characterized by the angles  $\alpha$ 
and $\beta$, respectively.
For the right piece, the angles at the same points are $\beta$ and $\alpha$, respectively. The equality 
$\alpha=\beta$ is obtained only when $\theta_{\FN}$=$\theta_{\ex}$.
Note that by antisymmetry $\psi_{\FN}(\theta+ \pi)=-\psi_{\FN}(\theta)$, the relative sign being arbitrary.

\begin{figure}[t]
\centering
  \includegraphics[width=0.9\columnwidth]{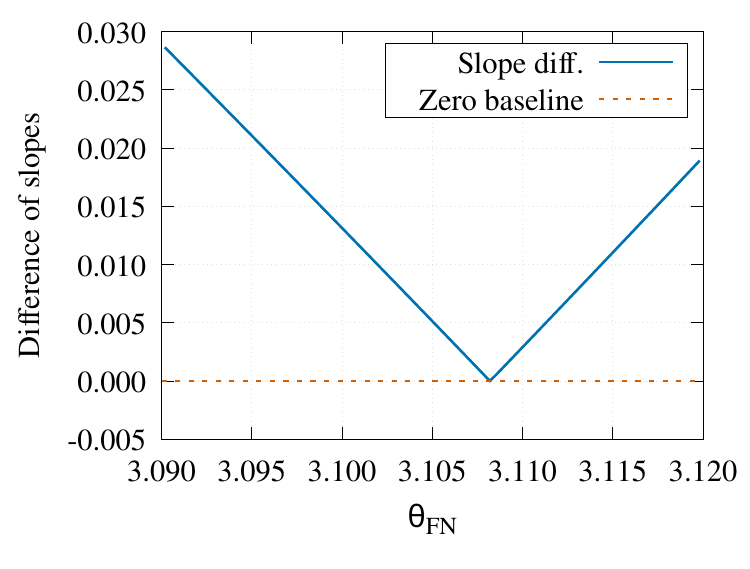}
\caption{Slope difference of the fixed-node wave function as a function of $\theta_{\FN}$.
        The difference vanishes at the exact node, $\theta_{\ex}$.}
\label{fig4}
\end{figure}

More quantitatively, the slope difference defined as
$\bigl|\psi^\prime_{\FN}(\theta^+_{\FN})-\psi^\prime_{\FN}(\theta^-_{\FN})\bigr|$
is plotted in Fig.~\ref{fig4} as a function of $\theta_{\FN}$. The difference vanishes at the exact node.

\section{The Self-Healing DMC algorithm}
\label{sec2}
\subsection{The SHDMC algorithm.}
As already mentioned, SHDMC is an iterative scheme that requires only fixed-node calculations. 
At each iteration, the nodal surface is updated using information obtained from the preceding fixed-node calculation. 
However, because the fixed-node calculations are performed using a Monte Carlo approach, great care must be 
taken to control the effect of the statistical noise, particularly near convergence of the nodal updates.
Previous authors have addressed this issue by introducing a filtering scheme designed to remove the noisiest components 
of the fixed-node wave function at each iteration (see, for example, \cite{Spanedda2025,Spanedda2026}).

To investigate the numerical convergence of the iterative process under the best possible conditions, 
it is desirable to eliminate the statistical fluctuations. For this reason, in the present study 
the fixed-node calculations are not performed with DMC, but instead in a (numerically) complete basis 
set subject to the fixed-node constraint. In this noise-free setting, the SHDMC algorithm can be written as follows.

The method starts from an approximate trial wave function $\psi_T^{(0)}$ expanded in a complete basis set
and truncated to the first $N$ basis functions. In our case,
the Fourier basis set $\{\phi_k\}$ introduced above is employed
\be
\psi_T^{(0)}(\theta) = \sum_{k=1}^{N} c^{(0)}_k\,\phi_k(\theta).
\ee
In the applications presented below,
the initial wave function is chosen as the variational solution in the truncated basis but other choices are possible.
Starting from $n=0$, the following steps are then iterated:

\begin{enumerate}

\item Locate the node $\theta_T^{(n)}$ of $\psi_{T}^{(n)}$ in the
interval $(0,\pi)$.

\item Perform a fixed-node calculation on the interval
	  $\bigl(\theta_{T}^{(n)},\,\theta_{T}^{(n)}+\pi \bigr)$ using the fixed-node basis set, $\{w_k\}$.
The number of basis functions is taken sufficiently large to ensure
the convergence of the fixed-node calculation. The fixed-node solution is denoted
$\psi_{\FN}^{(n+1)}$ with energy $E_{\FN}^{(n+1)}$.

\item Project the fixed-node solution back onto the $\{\phi_k\}$ basis set to give the new trial wave function
\be
\psi_{T}^{(n+1)}  = \sum_{k=1}^{N} c^{(n+1)}_k\,\phi_k(\theta),
\label{expansion}
\ee
where the updated coefficients are given by
\be
c^{(n+1)}_k = \left\langle \psi_{\FN}^{(n+1)} \Big| \phi_k \right\rangle,
\label{choice1}
\ee
that is,
\be
c^{(n+1)}_k =
\int_{\theta_T^{(n)}}^{\theta_T^{(n)}+\pi}
              \psi_{\FN}^{(n+1)}(\theta)
              \left[\phi_k(\theta)-\phi_k(\theta+\pi)\right]
            \,d\theta.
\label{choice2}
\ee

\item Return to step~1 and repeat until convergence.

\end{enumerate}

\begin{figure}[t]
\centering
  \includegraphics[width=0.9\columnwidth]{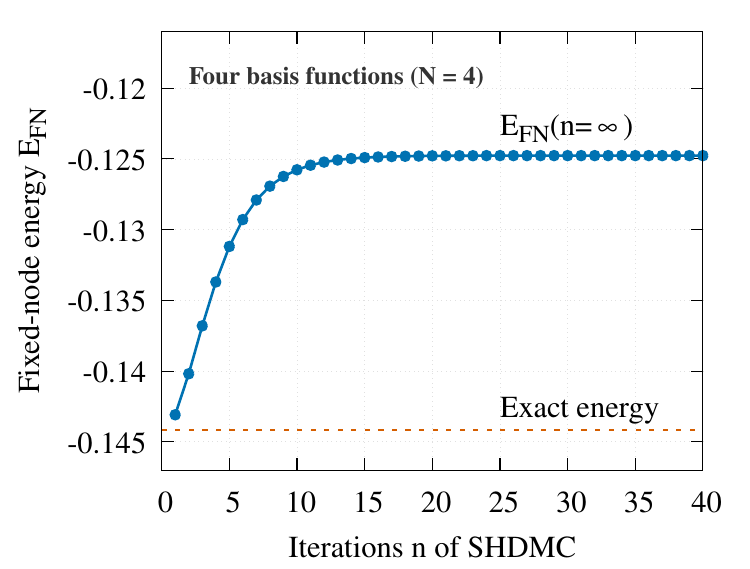}
	\caption{$E^{(n)}_{\FN}$ as a function of the number $n$ of SHDMC iterations.
The number of basis functions is $N=4$.}
\label{fig5}
\end{figure}

\begin{table}[t]
\centering
	\caption{Initial, $E^{(0)}_{\FN}$, and converged fixed-node energies, $E^{(\infty)}_{\FN}$,
	as a function of the number $N$ of basis functions.
	$\Delta E_{\FN}$ is the difference, $E^{(\infty)}_{\FN}-E^{(0)}_{\FN}$.}
        \label{tab1}
  \begin{tabular}{c@{\hspace{0.4cm}}c@{\hspace{0.4cm}}c@{\hspace{0.4cm}}c}
\hline
\hline
        $N$ &  $E^{(0)}_{\FN}$    &  $E^{(\infty)}_{\FN}$ & $\Delta E_{\FN}$ \\
\hline
        4 & $\num{-0.1430913460}$ & $\num{-0.124}$ &$\num{1.9E-002}$  \\
        6 & $\num{-0.1441429080}$ & $\num{-0.1434}$ &$\num{6.8E-004}$  \\
        8 & $\num{-0.1441518679}$ & $\num{-0.144148}$ &$\num{3.7E-006}$  \\
       10 & $\num{-0.1441519132}$ & $\num{-0.1441519115}$ &$\num{1.7E-009}$  \\
\hline
           $E_{\rm ex}=$&$\num{-0.1441519132}$  &      &     \\
\hline
\hline
\end{tabular}

\end{table}

Figure~\ref{fig5} shows the evolution of the fixed-node energy as a function of the number $n$ of SHDMC iterations
when expanding the trial wave function, $\psi_{T}^{(n)}$, over $N=4$ functions,
namely $\{ \phi_k \}_{N=4}=\{u_1,v_1,u_2,v_2\}$. 
The fixed-node energy increases monotonically
with $n$ and, for large $n$, approaches a plateau  corresponding to a biased energy.
This behavior is observed for all values of $N$ considered in this work, the quantitative results being presented in Table \ref{tab1}.

Table \ref{tab1} reports both the initial fixed-node energy, $E_{\FN}^{(0)}$, of SHDMC
and the corresponding converged value, as a function of the number $N$
of basis functions. For each value of $N$
the fixed-node energy converges to a biased value, and this bias decreases as the number of basis functions increases.
Remarkably, even for the large value $N=10$
where the initial fixed-node energy agrees with the exact energy to ten decimal places, $E_{\FN}=\num{-0.1441519132}$,
the SHDMC iterations still drive the energy upward, converging to a slightly biased value, $E_{\FN}=\num{-0.1441519115}$.
Taken together, these results show that, for this simple system and for any finite basis, SHDMC does not converge to the exact node:
starting from the variational solution, each iteration deteriorates the node, resulting in a monotonic increase of the fixed-node energy.

\begin{figure}[t]
\centering
  \includegraphics[width=0.9\columnwidth]{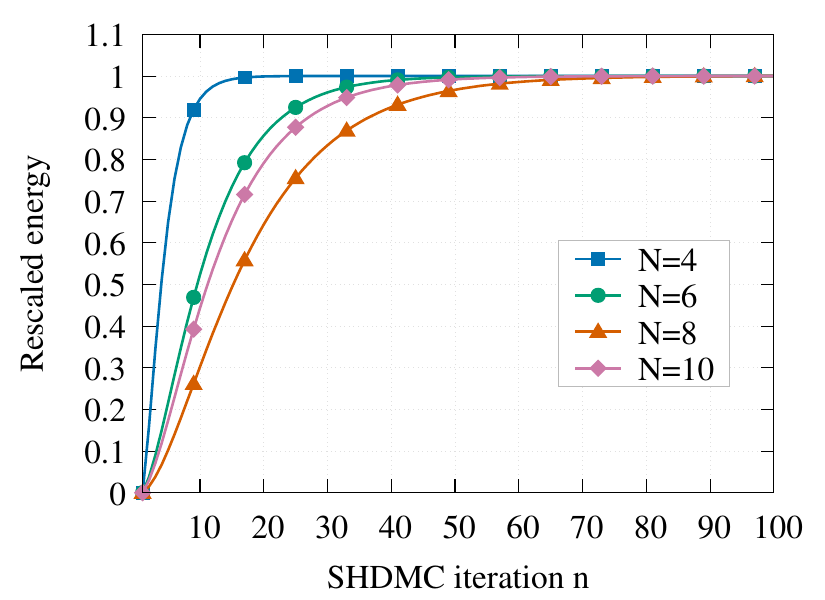}
	\caption{Comparison of the convergence of the fixed-node energies as a function of $n$ for $N=4,6,8$ and $10$.
         The uppermost curve corresponds to $N=4$, while the lower curves correspond to progressively larger basis sets.
	To facilitate the comparison, the energies
	have been rescaled according to, ${\tilde E}=(E^{(n)}_{\FN}-E^{(0)}_{\FN})/(E^{(100)}_{\FN}- E^{(0)}_{\FN})$.}
\label{fig6}
\end{figure}

Figure~\ref{fig6} compares the convergence curves obtained for different basis-set sizes, $N=4$, 6, 8, and 10.
To make a direct comparison on a single plot of very different energies,
the fixed-node energies have been rescaled to the interval $[0,1]$.
The uppermost curve corresponds to $N=4$, while the lower curves correspond to progressively larger basis sets.
Two main observations are in order. First, all curves reach a
plateau (equal to one by construction).
Second, the convergence rate decreases as the number of basis functions increases.
To understand the behavior observed above, let us take a closer look at the mathematical structure of the
SHDMC iteration process. More specifically, we focus on the node evolution under successive iterations.
Let $\theta_{T}^{(n)}$ denote the node at iteration $n$, and let $\Psi_{\FN}^{(n)}(\theta,\theta_{T}^{(n)})$ be
the associated fixed-node wave function, the dependence of the fixed-node wave function on $\theta_{T}$
being made explicit.
The optimal coefficients defining the updated trial wave function
are then obtained through the projection
\be
c^{(n+1)}_k = 2 \int_{\theta^{(n)}_{T}}^{\theta^{(n)}_{T}+\pi}  d\theta \Psi^{(n)}_{\FN}(\theta,\theta^{(n)}_{T})
\phi_k(\theta)
\ee
By exploiting the symmetry, the coefficients can also be obtained by integrating over the entire ring
\be
c^{(n+1)}_k = \int_0^{2\pi}  d\theta \Psi^{(n)}_{\FN}(\theta,\theta^{(n)}_{T}) \phi_k(\theta).
\label{cfull}
\ee
An important observation, which will play a central role in the following discussion, is that determining the "optimal" coefficients
by projecting the fixed-node solution onto the basis set [Eqs.~(\ref{choice1}) and (\ref{choice2})] is mathematically equivalent
to minimizing a ($\chi^2$-like) functional
with respect to these coefficients. Specifically, the corresponding $\chi^2$ functional is given by
\be
\chi^2 = \int_{0}^{2\pi}  d\theta \qty(\Psi^{(n)}_{\FN}(\theta,\theta^{(n)}_{T})-
\sum_{k=1}^N c_k \phi_k(\theta))^2.
\label{chi2}
\ee
The new node $\theta^{(n+1)}_{T}$ at iteration $n+1$ is expressed as a function of the preceding node using
the implicit equation
\be
\sum_{k=1}^N \qty[  \int_0^{2\pi}  d\theta \Psi^{(n)}_{\FN}(\theta,\theta^{(n)}_{T})
\phi_k(\theta) ] \phi_k(\theta^{(n+1)}_{T}) = 0.
\label{eqtheta}
\ee
Solving this equation iteratively defines a discrete-time dynamical system (a map) for the node, in which $n$
plays the role of time of evolution and $\theta$ some continuous variable. In a more general setting,
our equation reads
\be
f(x_n,x_{n+1})=0
\ee
where $f$ is the function resulting from Eq.(\ref{eqtheta}) (see also Eqs.~(\ref{fK}) and (\ref{KN}) below for a more transparent form).
The fixed points of the dynamical system obey
\be
f(x^{*},x^{*})=0.
\ee
Let us define the stability index $\eta$ as
\be
\eta \equiv \qty| \frac{f_x(x^{*},x^{*})}{f_y(x^{*},x^{*})}|
\ee
where $f_x$ and $f_y$ denote the partial derivatives of $f$ with respect to $x$ and $y$, respectively.
\\
\\
\begin{table}[H]
\centering
        \caption{Fixed points for each value of $N$. In each case the value of
	  the fixed-node angle and of the index $\eta$ are reported.}
        \label{tab2}
\begin{center}
\begin{tabular}{c cc cc}
\hline
\hline
$N$ & $\theta_N^*$ (repulsive) & $\eta$ & $\theta_N^*$ (attractive) & $\eta$ \\
4  & 1.438 & 1.35 & 0.184 \ ($\equiv 3.326$) & 0.67 \\
6  & 1.517 & 1.24 & 3.059 & 0.88 \\
8  & 1.524 & 1.16 & 3.109 & 0.90 \\
10 & 1.525 & 1.13 & 3.1059 & 0.92 \\
12 & 1.525 & 1.11 & 3.1057 & 0.93 \\
\hline
\hline
\end{tabular}
\end{center}
\end{table}

A fixed point is locally attractive if $\eta < 1$ and locally repulsive  if $\eta > 1$~\cite{Strogatz1994}.
In our case, $\eta$ is given by
\be
\eta =\qty|\frac{ \sum_{k=1}^N \qty[ \int d\theta \frac{ \partial \Psi_{\FN}}{\partial \theta_{T}}(\theta,\theta_{T})
\phi_k(\theta) ] \phi_k(\theta_{T})}
      { \sum_{k=1}^N \qty[ \int d\theta \Psi_{\FN}(\theta,\theta_{T}) \phi_k(\theta) ] \phi_k^{\prime}
      (\theta_{T})} |.
      \label{def_eta}
\ee
Table~\ref{tab2} reports the fixed points obtained for different values of $N$.
In each case, two fixed points are found: one attractive and one repulsive.
The corresponding values of $\eta$
are also given. Let us focus on the attractive fixed point, which governs the convergence of our calculations. The closer
$\eta$ ($<1$) is to unity, the slower the convergence toward this fixed point; in the limiting case
$\eta=1$, the convergence becomes infinitely slow. As can be seen,
 $\eta$  increases monotonically with $N$,
 thus explaining the trend observed in Fig.~\ref{fig6}.
Using Eq.~(\ref{cfull}), the function $f$ of Eq.~(\ref{eqtheta}) can be written, up to an irrelevant constant factor, as
\be
f(x,y) = \int_0^{2\pi} d\theta\, \Psi_{\FN}(\theta;x)\, K_N(\theta,y),
\label{fK}
\ee
where
\be
K_N(\theta,\theta') = \sum_{k=1}^{N} \phi_k(\theta)\phi_k(\theta')
\label{KN}
\ee
is the kernel associated with the projection onto the truncated basis. 
Using Eq.(\ref{def_eta}) and the completeness (resolution of the identity) at $N = \infty$ (within the odd subspace), that is
\be
\lim_{N\to\infty} K_N(\theta,\theta') = \frac{1}{2}\left[\delta(\theta-\theta') - \delta(\theta-\theta'-\pi)\right],
\label{delta}
\ee
where $\delta$ denotes the $2\pi$ periodic delta function on the ring,
it can be shown that the stability index $\eta$ tends to $1$ in the limit
$N \to \infty$. Indeed, Eq.~(\ref{fK}) gives
$f(x,y) \to \Psi_{\FN}(y,x)$ in this limit. This function vanishes at $y=x$ for any $x$, as a consequence of
the Dirichlet condition at the imposed node, Eq.~(\ref{dirichlet}). Differentiating $f(x,x)=0$ with respect to $x$ then gives
$f_x=-f_y$, that is, $\eta=1$. 
This result has a simple interpretation. In this limit, the
basis $\{\phi_k\}$ becomes complete, so the fixed-node solution is reproduced
exactly by its projection and the minimum of $\chi^2$ [Eq.~(\ref{chi2})]
vanishes for any nodal angle $\theta_T$. The node of the updated trial wave function therefore 
coincides with $\theta_T$: every angle is a fixed point of
the iteration, and SHDMC can no longer move the node.
This limiting behavior was already indicated by the authors of the method, who noted that,
with an infinite number of determinants and no statistical error, the trial wave function would exactly
reproduce the fixed-node solution, so that no iterative improvement of the nodes would be
possible~\cite{Reboredo2009a}. The analysis above provides a quantitative counterpart of this remark
and extends it to finite $N$.

\subsection{Relation to the original SHDMC algorithm}
\label{relation}

\paragraph*{Equivalence in the noise-free limit.}
In the original formulation~\cite{Reboredo2009a,Bajdich2010,Spanedda2026}, the trial wave function
is written as $\Psi_T = e^{J}\sum_n c_n \Phi_n$, where the $\Phi_n$ form an orthonormal antisymmetric basis, $e^{J}$ a symmetric 
Jastrow factor,
and the coefficients are updated as $c_n = \langle \Phi_n | e^{-J} | \Psi_0 \rangle$, $\Psi_0$ being the
fixed-node solution. These coefficients are estimated from the DMC mixed distribution, $\psi_T \Psi_0$.
For $J=0$ and an infinite number of sampled configurations, this update reduces exactly to
Eq.~(\ref{choice1}). Ordering our basis functions by increasing kinetic
energy ({\it i.e.} increasing $k's$ in Eq.(\ref{Fourier_basis})) and truncating the expansion, Eq.(\ref{expansion}), corresponds to the energy-cutoff truncation discussed in
Ref.~\cite{Reboredo2009a}. The scheme studied here is therefore the deterministic limit of SHDMC.

\paragraph*{Adaptive truncation.}
In practical implementations, an active space of determinants is chosen before the iterations begin:
for instance, the full configuration-interaction basis of a model system ordered by non-interacting
energy~\cite{Reboredo2009a}, the most important excitations of a configuration-interaction
expansion~\cite{Bajdich2010}, or a subset of a selected configuration-interaction
expansion~\cite{Spanedda2025}. All the coefficients of the active space are sampled at each iteration, but only
those that are statistically significant (typically $|\langle c_n \rangle| \ge 4\sigma_{c_n}$, where $\sigma_{c_n}$ is the estimated
standard error on the coefficient) enter the trial wave function of the next iteration; the others are set to zero, but continue to be
sampled~\cite{Spanedda2025,Spanedda2026}. The number of samples per iteration is never decreased; it is increased
(by a factor 1.4 in Refs.~\cite{Spanedda2025,Spanedda2026}) whenever two successive updates of the coefficient vector
point in opposite directions, a signature of dominant statistical noise. As the standard error decreases, more and more
coefficients pass the signal-to-noise cutoff,
so that the number of retained determinants grows along the
iterations, the most significant ones entering first. In addition, the active space itself may be enlarged, either
manually~\cite{Spanedda2025} or, in the ``auto-branching'' algorithm of Ref.~\cite{Spanedda2026}, by adding the
particle--hole excitations of the significant single and double excitations, subject to a mean-field energy cutoff.
Branching is triggered when the fluctuation criterion above is met and the DMC energy increases from one
iteration to the next.

Consider now the limit of infinite statistics at each iteration. All the standard errors vanish, so that every
nonzero coefficient passes the signal-to-noise criterion from the first iteration on: the whole active space is
retained immediately, and there is no progressive growth. The fluctuation criterion, designed to detect noise, is
not met for a monotonically converging deterministic iteration, and the auto-branching procedure is therefore not
triggered either. The practical algorithm then reduces exactly to the fixed-$N$ scheme studied here, with $N$ equal
to the size of the active space. Furthermore, since the statistics increase along the iterations, all the
significant coefficients are eventually retained. When the iteration possesses a single attractive fixed point, as
in the present model, the progressive growth of the number of retained determinants therefore affects the path and
the rate of convergence, but not the end point, which is the fixed point of the deterministic map restricted to the
retained determinants. The only difference comes from coefficients that never exceed the signal-to-noise cutoff,
so that the final effective basis may be somewhat smaller than the active space. In this sense, the noise-free
scheme describes not only the infinite-statistics limit of SHDMC, but also, to a good approximation, the end point
of its practical implementations.

\paragraph*{Role of statistical noise.}
The original authors argued that statistical fluctuations of the coefficients play a beneficial role,
akin to a temperature in simulated annealing, which allows the nodes to move~\cite{Reboredo2009a}.
The noise-free map studied here describes the mean drift of the iteration: to leading order in the noise
amplitude, the stationary distribution of the node is centred on the attractive fixed point of the
deterministic map. Whether statistical noise combined with the filtering procedure can induce a systematic
drift toward the exact node beyond this deterministic behavior is not addressed in the present work.

\paragraph*{Theoretical justification of SHDMC.}
The theoretical basis of SHDMC is the observation that smoothing the kink of the fixed-node wave function
by convolution with an approximation of the Dirac delta function moves the node toward the exact
one~\cite{Reboredo2009a}. The corresponding displacement is proportional to the width $\sqrt{\tau}$
of the smoothing kernel, and the argument assumes a kernel that is local on the scale of the variations
of the nodal surface and has a positive weight. In the present case, the smoothing kernel implied by
Eqs.~(\ref{choice1}) and (\ref{fK}) is the kernel $K_N$ of Eq.~(\ref{KN}), which is 
a translation-invariant antisymmetrized delta function, Eq.(\ref{delta}).
However, for small $N$ this kernel is broad. Writing $u=\theta-\theta'$, the kernel takes the expression
$
K_N(\theta,\theta') \sim \frac{\sin(N u)}{2\pi \sin u},
$
so that its central lobe has a half-width $\pi/N$, that is, $\pi/4$ for $N=4$. This is a sizeable
fraction of the nodal domain, of length $\pi$, and of the scale of variation of the potential, the
$\sin 4\theta$ term having period $\pi/2$. The small-$N$ regime therefore does not satisfy the locality
assumption on which the original argument relies\cite{Reboredo2009a}.
Conversely, for large $N$ the kernel becomes local,
the attractive fixed point approaches the exact node (Table~\ref{tab2}), and the displacement per
iteration vanishes ($\eta \to 1$), consistently with a displacement proportional to the kernel width.
Our results thus do not contradict the original argument; they delimit its range of validity.
Note also that the kernel above is built from eigenfunctions of the kinetic operator only and
contains no information on the potential, whereas in practical applications the determinants are built
from mean-field orbitals that incorporate a large part of it. 
In the extreme case where the basis consists of eigenfunctions of $H$ itself, $\psi_1$
is one of the basis functions and the exact node is trivially a fixed point.

\paragraph*{Best node attainable in a given basis.}
It has been stated that repeated application of SHDMC yields the best nodal surface for a given
basis~\cite{Bajdich2010}. The present model provides a counterexample. Since the node consists of a single
parameter, the exact node can be represented with any of the bases used here (for instance,
$\sin(\theta-\theta_{\ex}) = \sqrt{\pi}\,(\cos\theta_{\ex}\, v_1 - \sin\theta_{\ex}\, u_1)$).
Since the fixed-node energy depends only on the position of the node, and not on the rest of the
trial wave function, the exact fixed-node energy $E_{\ex}$ is therefore attainable within each of the bases
of Table~\ref{tab1}: the best node attainable in a given basis is the exact one. Yet SHDMC
converges to $\theta^*_N \ne \theta_{\ex}$; for $N=4$, the converged node, $3.326$, differs from
$\theta_{\ex}=3.1058$ by $0.22$~rad, which is further than the trivial symmetric guess $\theta=\pi$.
The reason is that SHDMC does not optimize the node itself, but a global fit of the wave function,
Eq.~(\ref{chi2}). When $\theta_T=\theta_{\ex}$, the fixed-node solution is $\psi_1$; 
if $\psi_1$ can be expanded exactly in the basis, its projection is $\psi_1$
itself
and the exact node is a fixed point. Otherwise,
the projection of $\psi_1$ generally does not vanish at $\theta_{\ex}$. In other words, the representability
of the exact node does not imply the representability of the exact wave function, and it is the latter
that controls the SHDMC fixed point.

\paragraph*{Dependence on the initial trial wave function.}
The increase of the fixed-node energy along the iterations, observed in Fig.~\ref{fig5} for $N=4$ and
in Table~\ref{tab1} for all values of $N$ ($\Delta E_{\FN}>0$), results from the choice of the variational
solution as starting point, whose fixed-node energy is lower than the fixed-node energy at the attractive
fixed point $\theta^*_N$. When the iterations are started from a poorer node, they
converge to the same attractive fixed point, but with a decreasing energy. This is expected, since the map
has only two fixed points in $(0,\pi)$, the other one being repulsive. The converged result is thus
independent of the initial trial wave function, in agreement with the observations reported in
Refs.~\cite{Bajdich2010,Spanedda2025}, but the common limit is biased.

\paragraph*{Relevance of the finite-basis bias for realistic systems.}
The bias decreases rapidly with $N$ (Table~\ref{tab1}): for $N=10$, it is only $1.7\times10^{-9}$,
far below any realistic statistical resolution. In this model, however, the node reduces to a single point
on the ring, determined by one parameter, whereas in realistic systems it is a hypersurface of dimension
$3N_e-1$ in the configuration space of the $N_e$ electrons. Describing such a surface accurately would
require a very large number of determinants, so that the expansions used in practice are far from being complete.
We therefore believe that the regime of the model most relevant to realistic applications is the small-$N$
one, in which the basis is small compared with what an exact description of the node would require.
Quantifying this statement for realistic systems, however, remains an open question.

\subsection{SHDMC focusing on the nodal region}
\label{sec3}
As seen in the previous section, SHDMC applied to our model with a finite basis fails to improve the node and, worse,
appears to systematically deteriorate  the initial nodal position, converging instead to an incorrect node.
To address this, it is natural to modify the criterion used in the optimization of the coefficients
so as to give more weight to the nodal region. We therefore propose restricting the integration interval in the
$\chi^2$ functional, Eq.(\ref{chi2}), to a region close to the node.

The new functional is
\be
\chi^2_\epsilon= \int_{I_\epsilon} d\theta \qty( \psi^{(n)}_{\FN} -  \sum_{k=1}^{N} c^{(n)}_k \phi_k(\theta))^2
\ee
where the integration is now performed in the interval $I_\epsilon$
\be
I_\epsilon = [\theta^{(n)}_{T}-\epsilon,\theta^{(n)}_{T}+\epsilon] \cup [\theta^{(n)}_{T} +\pi-\epsilon,
\theta^{(n)}_{T}+\pi+\epsilon]
\ee
where $\epsilon$ can be varied from 0 to $\frac{\pi}{2}$.
The minimization of $\chi^2_\epsilon$ with respect to the coefficients gives
\be
c^{(n)} = S^{-1} u^{(n)}
\label{cS}
\ee
where the partial overlap matrix $S$ is given by
\be
S_{ij} = \int_{I_\epsilon}  d\theta\, \phi_i(\theta) \phi_j(\theta)
\ee
and the components of the vector $u^{(n)}$
\be
u^{(n)}_k = \int_{I_\epsilon}  d\theta\, \psi^{(n)}_{\FN}(\theta) \phi_k(\theta).
\ee
Using the inversion symmetry, we get
\begin{align}
  u^{(n)}_k  = & \int_{\theta^{(n)}_{T} + \pi - \epsilon}^{\theta^{(n)}_{T} +\pi } d\theta\, \psi^{(n)}_{\FN}(\theta) \qty(  \phi_k(\theta) - \phi_k(\theta-\pi)) \nonumber \\
  & + \int_{\theta^{(n)}_{T}}^{\theta^{(n)}_{T} + \epsilon}  d\theta\, \psi^{(n)}_{\FN}(\theta) \qty( \phi_k(\theta) - \phi_k(\theta+\pi))
\end{align}

\begin{figure}[t]
\centering
  \includegraphics[width=0.9\columnwidth]{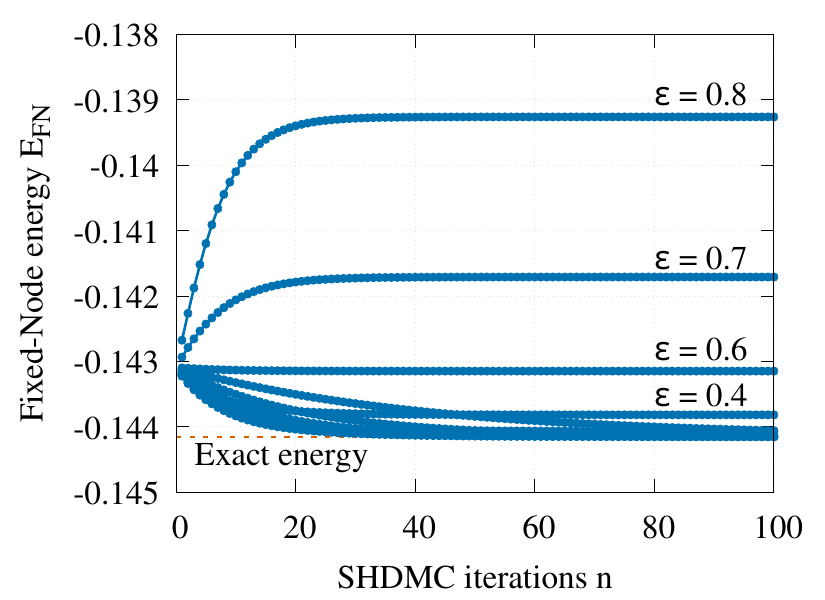}
\caption{$N=4$. Convergence of the SHDMC fixed-node energy as a function of SHDMC iterations. From top to bottom,
curves correspond to decreasing values of $\epsilon$,
	ranging from 0.8 to 0.05.}
\label{fig7}
\end{figure}

\begin{figure}[t]
\centering
  \includegraphics[width=0.9\columnwidth]{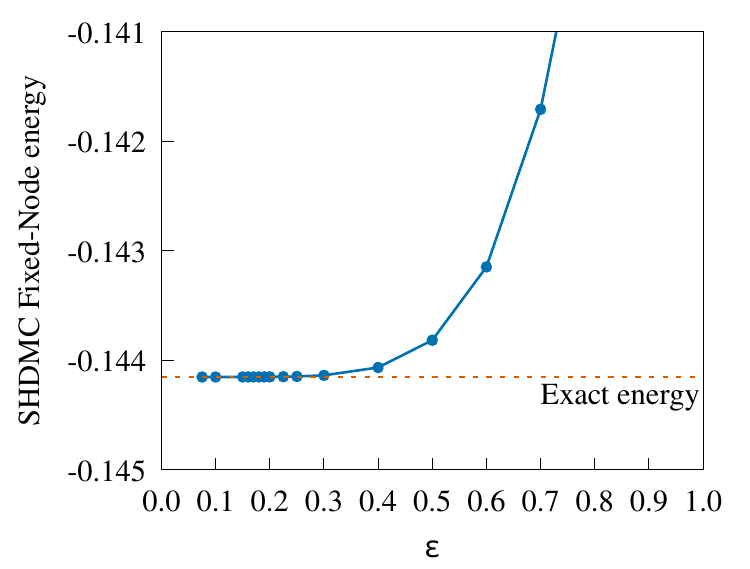}
        \caption{$N=4$. Converged SHDMC fixed-node energy as a function of $\epsilon$.}
\label{fig8}
\end{figure}

\begin{figure}[t]
\centering
  \includegraphics[width=0.9\columnwidth]{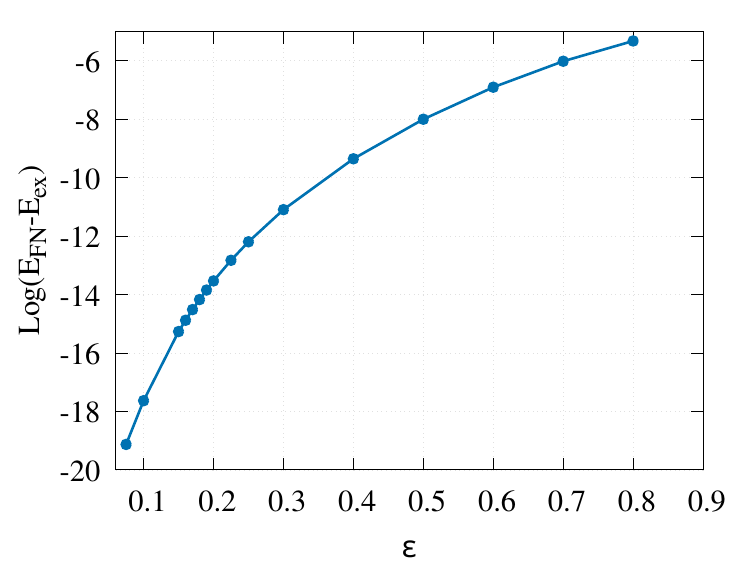}
        \caption{$N=4$. Logarithm of the energy bias as a function of $\epsilon$}
\label{fig9}
\end{figure}

In Fig.~\ref{fig7}, the convergence curve of the SHDMC fixed-node energy as a function of the iterations is plotted
for different values of $\epsilon$ and $N=4$.
As seen, for each value of  $\epsilon$,
the fixed-node energy converges to a biased value; however, two distinct regimes are observed.
For large values of  $\epsilon$, the fixed-node energy increases upon iteration,
as was the case in the previous section, thus deteriorating the
node.
Below a critical value of $\epsilon \sim 0.6$,
however, the energy instead decreases during the SHDMC iterations,
indicating that the node is now improved by the iterative algorithm.
The converged fixed-node energy as a function of  $\epsilon$
is reported in Fig.~\ref{fig8}. For $\epsilon \lesssim 0.3$,
the asymptotic energies are virtually indistinguishable from the exact value, whereas the accuracy gradually deteriorates as
$\epsilon$
increases. From this plot alone,
it is difficult to determine whether the fixed-node energy remains exactly equal to the exact energy over a {\it finite} range of small
$\epsilon$ and then increases, or whether a (small) bias is already present as soon as $\epsilon >0$.
To investigate this point, Fig.~\ref{fig9} displays the logarithm of the energy difference
\be
\Delta E = E_{\FN}^{(\infty)}-E_{\rm ex}
\ee
as a function of $\epsilon$.
The figure clearly shows that the bias $\Delta E$
is nonzero for all $\epsilon$.
A fit of the data suggests that the behavior at small $\epsilon$
is compatible with
\be
\Delta E \sim \exp\left(-\frac{a}{\epsilon^\alpha}\right),
\ee
with an exponent in the range $\alpha \approx 0.4$--$0.5$ (for $\alpha=0.5$,  $a \sim 6$).
If this form holds,  $\Delta E$
is a non-analytic function at $\epsilon=0$: All derivatives vanish at the origin,
resulting in the extremely flat behavior at small $\epsilon$
observed in Fig.~\ref{fig8}.

Two remarks are in order regarding the use of this criterion in realistic DMC calculations.
First, the density of walkers, $f=\Psi_T\Psi_{\FN}$, vanishes quadratically at the nodes, so that few walkers
sample $I_\epsilon$ when $\epsilon$ is small, while the estimators of the coefficients, which involve the ratio
$\Phi_n/\Psi_T$, fluctuate strongly in the nodal region.
Estimating the vector $u^{(n)}$ with a given accuracy is therefore expected to
become increasingly costly as $\epsilon$ decreases. Second, the partial overlap matrix $S$ becomes
ill-conditioned as $\epsilon \to 0$, since the restrictions of the basis functions to $I_\epsilon$ become
nearly linearly dependent. 

\subsection{Use of a localized basis set}
\label{local}
In the previous section, we showed that a substantially improved node can be obtained
by employing an optimization criterion
that gives greater weight to the nodal region.
However, using such a criterion may significantly deteriorate the quality of
the optimized trial wave function away from the nodes.
While this is not problematic in a purely deterministic framework,
it becomes a serious issue in Diffusion Monte Carlo, where the trial wave function must remain accurate in the regions of
highest probability density in order to keep statistical errors small.
In other words, one would like to improve the node while preserving the quality of the wave function in the physically important regions
of configuration space. A natural way to achieve this goal is to employ a localized basis set, which allows the node to be moved through
a local distortion of the wave function.

To construct such a basis, we introduce a uniform grid of $N$ angles on the interval $(0,\pi)$,
\be
\theta_k = k\Delta\theta , \qquad k=0,\ldots,N-1,
\ee
with
\be
\Delta\theta=\frac{\pi}{N}.
\ee
The nonorthogonal localized basis set functions are now defined by
\be
\phi_k(\theta)= \sum_{l=-\infty}^{\infty} \qty[e^{-\gamma \qty( \theta-\theta_k + 2\pi l)^2} - e^{-\gamma \qty( \theta-(\theta_k+\pi) + 2\pi l)^2} ].
\ee
These basis functions satisfy
\be
\phi_k(\theta+2\pi)=\phi_k(\theta),
\qquad
\phi_k(\theta+\pi)=-\phi_k(\theta),
\ee
and are, therefore, consistent with the symmetry of the exact fermionic solution.
Since this basis is nonorthogonal, the coefficients are obtained through the overlap matrix,
Eq.~(\ref{cS}) with the integration extended to the full ring for the usual criterion, 
which is equivalent to an orthogonal projection onto the space spanned by the basis functions.

\begin{figure}[t]
\centering
  \includegraphics[width=0.9\columnwidth]{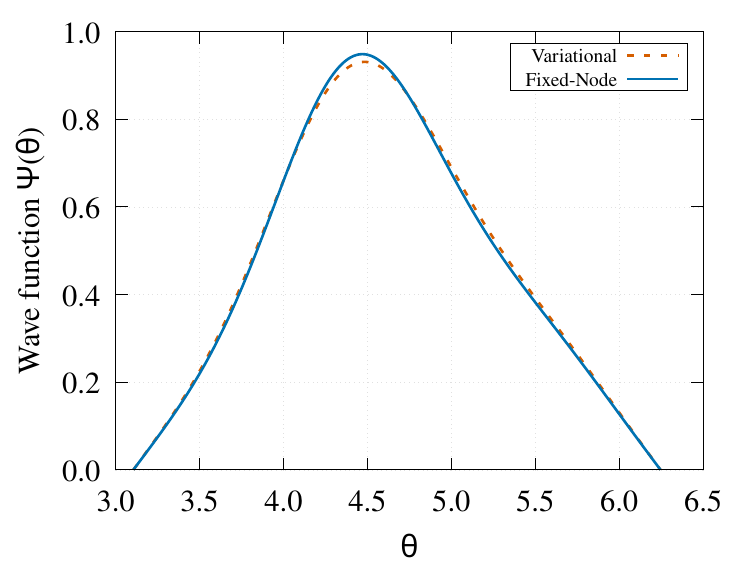}
	\caption{$N=5$. Use of a localized basis set. Comparison between the optimized  fixed-node (solid line) and
	the distorted variational (dashed) wave functions.
        The resulting node of the optimized FN wave function is at $\num{3.1051}$, compared with the exact value of $\num{3.1058}$.}
	\label{fig10}
\end{figure}

When the usual SHDMC criterion is used, the behavior is qualitatively similar to that observed
with the delocalized Fourier basis: the algorithm converges, but toward a fixed-node solution
with a biased energy. In contrast, when the new local criterion is employed,
physically meaningful solutions can be obtained for moderately small values of $\epsilon$.
In this regime, the algorithm converges to a very accurate node while maintaining a trial wave function of satisfactory quality.
To give a quantitative example, we consider the case $N=5$ with $\gamma=2$.
The iterative process is initiated with a variational wave function of energy $E_{\rm var}=\num{-0.1439225}$.
At $\epsilon=0.5$, the algorithm converges to the very accurate fixed-node energy of
$E_{\FN}=\num{-0.14415176}$
compared with the exact energy of $\num{-0.14415191}$.
The variational energy of the distorted trial wave function with optimized SHDMC coefficients deteriorates to
$E_{\rm var}=\num{-0.1423031}$, that is, the variational error increases from $\num{2.3E-4}$ to $\num{1.85E-3}$.
This example illustrates the trade-off involved: the nodal error is reduced to a fixed-node energy error
of about $\num{1.5E-7}$, while the overall quality of the trial wave function deteriorates moderately but
significantly. The trial wave function nevertheless remains of reasonable quality over the full angular
range, as illustrated in Fig.~\ref{fig10}, where the distorted variational wave function and the optimized
fixed-node solution are plotted as a function of the angle: the two are in good qualitative agreement.
In DMC, however, even a moderate deterioration of the trial wave function increases the variance of the
local energy, and hence the statistical error. In the present model, the trade-off between nodal accuracy
and overall quality of the trial wave function is easily controlled, since a local distortion of the wave
function is sufficient to move the single nodal point. It is not clear that such a favorable trade-off can be
achieved in realistic systems, where the node is a high-dimensional hypersurface and where the
construction of a basis allowing local distortions of the wave function is far less straightforward.

\section{Summary}
\label{sec4}
In this work, we have analyzed the SHDMC algorithm using a simple one-dimensional model that captures
the essential features of fermionic systems. Our main result is that, in its noise-free limit, the SHDMC
iteration converges, for any \emph{finite} basis, to an attractive fixed point that differs from the exact
node, even though the exact node can be represented in all the bases considered. The SHDMC fixed point is
therefore not, in general, the best node attainable in a given basis. This fixed point approaches the
exact node as the size $N$ of the basis increases, but the displacement of the node at each iteration then
becomes vanishingly small, so that the convergence becomes increasingly slow. SHDMC thus faces a dilemma:
with a small basis, the node converges quickly but to a biased value; with a large basis, the converged
node is accurate, but it is reached only after a very large number of iterations.

This dilemma raises a question for realistic electronic systems, in which the trial wave function is
expanded over a set of Slater determinants. SHDMC has been applied to such systems, with
significant improvements of the
fixed-node energies: 
for atoms and small molecules, the SHDMC coefficients and energies were found to agree with those
of large configuration-interaction expansions and of variational energy minimization~\cite{Bajdich2010},
and for periodic solids, the converged energies and wave functions were found to be independent of the
initial trial wave function and close to complete-basis-set selected configuration-interaction
estimates~\cite{Spanedda2025,Spanedda2026}. 
These comparisons, however, are made in the presence of statistical uncertainties
that are generally
significantly larger than the biases discussed here. In addition, in the large-$N$ regime, where the
displacement of the node per iteration is very small, a slow systematic drift of the node may be hidden 
by the statistical fluctuations of the coefficients, so that an apparent convergence of the iterations
does not necessarily reflect true convergence.
Moreover, as shown in Sec.~\ref{relation}, independence with respect
to the initial guess does not guarantee convergence to the exact node. Since the nodal hypersurface of a
many-electron system is a complex object, whose accurate description certainly requires a large number of
determinants, a residual bias of the type identified in this work cannot be excluded. 

Nevertheless, the present model offers valuable insight into the optimization process. In particular,
it suggests that the criterion used to optimize the coefficients of the trial wave function should place
more emphasis on the nodal region. The driving force toward the exact nodes is the reduction, as much as
possible, of the discontinuity in the derivative at the node. An effective optimization strategy should
therefore allow local distortions of the wave function in the vicinity of the nodes, while minimally
affecting the high-probability regions located far from them. From this perspective, the use of localized
basis functions capable of producing such local deformations appears to be an important ingredient of a
successful SHDMC implementation.

The extension of these conclusions to many-electron systems is not straightforward. In particular, what
constitutes a ``localized'' basis in the space of Slater determinants, and whether statistical noise
combined with the signal-to-noise filter used in practical implementations can modify the conclusions
drawn from the noise-free map, remain open questions. Natural extensions of the present study include
(i) the addition of some artificial noise and of the signal-to-noise filter to the deterministic scheme, which
would directly address the role of statistical fluctuations, and (ii) the use of a basis of eigenfunctions
of an approximate Hamiltonian (for instance, $\lambda=0$), mimicking mean-field orbitals.

{\it Acknowledgments}. We are grateful to Dr. Laurent Miclo (Toulouse School of Economics)
for bringing the model studied in this work to our attention.
We are also grateful to Dr. F. A. Reboredo for exchanges on the SHDMC method, 
which motivated us to clarify the relation between the present noise-free scheme and the original algorithm. 
\appendix

\section{Free particle on a segment}
Let us consider a free particle confined to the segment $[-\frac{1}{2},\frac{1}{2}]$. The eigenfunctions are given by
\begin{equation}
\psi_n(x) = \sqrt{2}\, \sin{\qty[n\pi\qty(x+\tfrac{1}{2})]}, \qquad n=1,2,\ldots
\label{eq:eigenfunctions}
\end{equation}
By symmetry, these eigenfunctions are either even or odd. The corresponding energies are
\begin{equation}
E_n = \frac{n^2\pi^2}{2}.
\label{eq:energies}
\end{equation}
The ground state, $n=1$, has no node except at the boundaries and plays the role of the ``bosonic'' ground state.
The first excited state, $n=2$, has its node at 0 and plays the role of the ``fermionic'' ground state;
here the location of the node is fixed by symmetry, which makes this simple system
not representative of realistic fermionic systems, but yet provides a useful test case
for which the equations can be worked out analytically.

Let us now consider a trial wave function with an approximate node located at $x_{\FN}$. For $x_{\FN}>0$, the fixed-node wave function on the left interval $[-\frac12,x_{\FN}]$ is
\begin{equation}
\Psi^L_{\FN}(x) = \frac{1}{\sqrt{x_{\FN}+\frac12}}\, \sin{\qty(\frac{\pi\qty(x+\frac12)}{x_{\FN}+\frac12})},
\label{eq:psiL}
\end{equation}
and, on the right interval $[x_{\FN},\frac12]$,
\begin{equation}
\Psi^R_{\FN}(x) = \frac{1}{\sqrt{\frac12 - x_{\FN}}}\, \sin{\qty(\frac{\pi(x-x_{\FN})}{\frac12-x_{\FN}})}.
\label{eq:psiR}
\end{equation}
Both functions are normalized to $\frac12$. The absolute values of their derivatives at the node are
\begin{equation}
\qty|\Psi^{\prime}_{\FN}(x_{\FN}^-)| = \frac{\pi}{\qty(x_{\FN}+\frac12)^{3/2}},
\end{equation}
\begin{equation}
\qty|\Psi^{\prime}_{\FN}(x_{\FN}^+)| = \frac{\pi}{\qty(\frac12 - x_{\FN})^{3/2}},
\end{equation}
which coincide when the node is exact, $x_{\FN}=0$.

As a trial function, we use a polynomial form,
\begin{equation}
\Psi_T(x) = \sum_{k=1}^{N} c_k\, u_k(x),
\end{equation}
with basis functions $u_k$ constrained to vanish at the boundaries,
\begin{equation}
u_k(x) = \qty(x+\frac12)\qty(x-\frac12)\, x^{k-1}.
\end{equation}

\begin{figure}[b]
\centering
  \includegraphics[width=0.9\columnwidth]{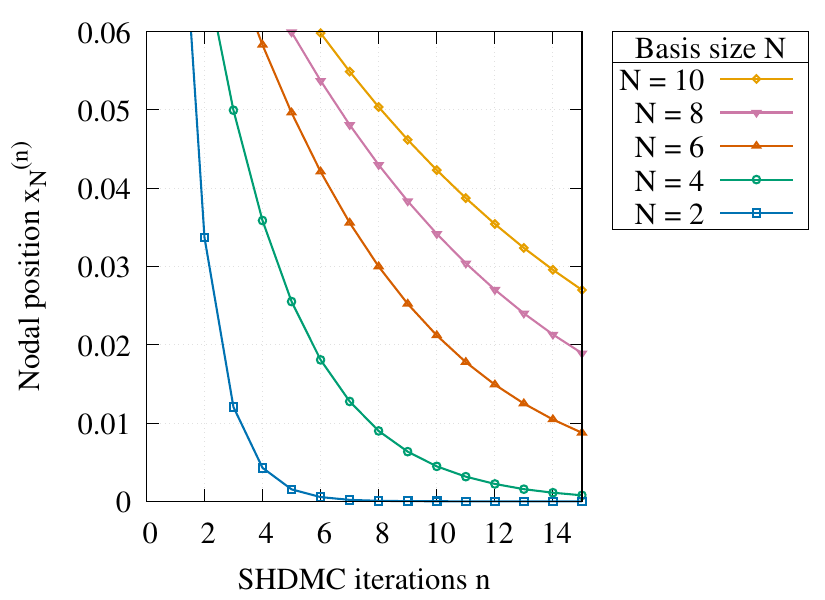}
\caption{Convergence of $x_N$ as a function of the SHDMC iterations, for different numbers of basis functions $N$.}
\label{fig11}
\end{figure}

At iteration $n$, the node is denoted $x_{\FN}^{(n)}$ and the trial coefficients $c_k^{(n)}$. They evolve according to
\begin{equation}
c_k^{(n+1)} = \sum_{l=1}^N S^{-1}_{kl} \langle \Psi_{\FN}^{(n)} | u_l \rangle,
\end{equation}
where the overlap matrix is
\begin{equation}
S_{kl} = \langle u_k|u_l\rangle = \int_{-1/2}^{1/2} dx\, \qty(x^2-\tfrac14)^2\, x^{k+l-2},
\end{equation}
that is,
\begin{equation}
S_{kl} = \frac{1+(-1)^{k+l}}{ 2^{k+l}(k+l+3)(k+l+1)(k+l-1)}.
\end{equation}
We also have
\begin{widetext}
\begin{equation}
\langle \Psi_{\FN}^{(n)}|u_l\rangle = \frac{1}{\sqrt{x_{\FN}^{(n)}+\frac12}} \int_{-1/2}^{x_{\FN}^{(n)}} dx\, \sin{\qty(\frac{\pi(x+\frac12)}{x_{\FN}^{(n)}+\frac12})} u_l(x) - \frac{1}{\sqrt{\frac12-x_{\FN}^{(n)}}} \int_{x_{\FN}^{(n)}}^{1/2} dx\, \sin{\qty(\frac{\pi(x-x_{\FN}^{(n)})}{\frac12-x_{\FN}^{(n)}})} u_l(x),
\end{equation}
\end{widetext}
where we have chosen a minus sign in front of the right fixed-node contribution.

Figure~\ref{fig11} shows the convergence of the node with the number of iterations, for several values of $N$. In every case the node converges 
to its exact value of 0, a result which shows that for this specific model the SHDMC method
does work. Interestingly, the convergence rate to the exact node decreases as $N$ increases, as was also observed 
for the ring model.

Let us make the equations explicit for $N=2$. At iteration $n$,
\begin{widetext}
\begin{align}
\langle \Psi_{\FN}^{(n)}|u_1\rangle &\equiv A^{(n)} = 
\frac{\qty(x_{\FN}^{(n)}+\frac12)^{3/2}}{\pi^3}\qty[-4\qty(x_{\FN}^{(n)}+\frac12)+\pi^2\qty(x_{\FN}^{(n)}-\frac12)] \notag\\
&\quad + \frac{\qty(\frac12-x_{\FN}^{(n)})^{3/2}}{\pi^3}\qty[4\qty(\frac12-x_{\FN}^{(n)})+\pi^2\qty(x_{\FN}^{(n)}+\frac12)],
\end{align}
\begin{align}
\langle \Psi_{\FN}^{(n)}|u_2\rangle &\equiv B^{(n)} = \frac{\qty(x_{\FN}^{(n)}+\frac12)^{3/2}\qty(x_{\FN}^{(n)}-\frac12)\qty(-3+(\pi^2-6)\,x_{\FN}^{(n)})}{\pi^3} \notag\\
&\quad + \frac{\qty(\frac12-x_{\FN}^{(n)})^{3/2}\qty(x_{\FN}^{(n)}+\frac12)\qty(3+(\pi^2-6)\,x_{\FN}^{(n)})}{\pi^3}.
\end{align}
\end{widetext}

Since the overlap matrix $S$ is diagonal for $N=2$, the coefficients evolve as
\begin{equation}
c_1^{(n+1)} = S^{-1}_{11}\, A^{(n)}, \qquad c_2^{(n+1)} = S^{-1}_{22}\, B^{(n)}.
\end{equation}
The node of the trial wave function is given by $x_{\FN}^{(n+1)} = -c_1^{(n+1)}/c_2^{(n+1)}$, {\it i.e.},
\begin{equation}
x_{\FN}^{(n+1)} = -\frac{S^{-1}_{11}\, A^{(n)}\qty[x_{\FN}^{(n)}]}{S^{-1}_{22}\, B^{(n)}\qty[x_{\FN}^{(n)}]},
\end{equation}
which expresses $x_{\FN}^{(n+1)}$ as a function of $x_{\FN}^{(n)}$. 
By linearizing the dynamics in the vicinity of the fixed point $x^{*}=0$, we get
\begin{equation}
x_{\FN}^{(n+1)} = \alpha\, x_{\FN}^{(n)},
\end{equation}
with $\alpha$ a positive constant given by
\begin{equation}
\alpha = \frac{20+\pi^2}{84} \sim 0.356
\end{equation}
This confirms the geometric convergence of the iterative process close to the fixed point.

\end{document}